\documentstyle[prb,aps,preprint]{revtex}
\begin{document}
\draft
%
%
\title{Infrared Spectroscopic Study of Phonons Coupled to Charge Excitations in FeSi}
\author{A. Damascelli, K. Schulte, D. van der Marel,}
\address{Solid State Physics Laboratory, University of Groningen, 
Nijenborgh 4, 9747 AG Groningen, The Netherlands}
\author{A. A. Menovsky} 
\address{Van der Waals-Zeeman Laboratory, University of Amsterdam, 
Valckenierstraat 67,  1018 XE Amsterdam, The Netherlands} 
\date{\today}
\maketitle
\begin{abstract}
From an investigation of the optical conductivity of FeSi single crystals 
using FTIR spectroscopy in the frequency range from 30 to $20\,000$ cm$^{-1}$ 
we conclude that the transverse effective charge of the Fe and Si ions is
approximately $4e$. Of the five optical
phonons which are allowed by symmetry we observe only four, three of which
have a Fano line shape presumably resulting from an interaction of these 
modes with the electronic continuum. We show  that
the large oscillator strength of the phonons results from a relatively weak
coupling  ($\lambda \approx 0.1$) of the lattice degrees of freedom to 
an electronic resonance above the semiconductor gap, 
which is also responsible for the large electronic polarizability 
($\epsilon_{\infty} \approx 200$) of the medium.
\end{abstract}
%
%
\vskip2pc
\narrowtext
FeSi is a fascinating material which has been studied 
for many years for its unusual 
magnetic and thermodynamical properties.\cite{jaccarino}  
To explain the anomalies in the magnetic 
susceptibility,\cite{jaccarino}  thermodynamic\cite{mandrus} and 
spectroscopic properties\cite{schlesinger,park}
several models have been proposed involving strong 
electron-electron correlations.\cite{aeppli,mattheiss,fu,tmrice}  

In this paper we present a detailed analysis of the 
sharp absorption lines observed in the far 
infrared region. First we solve an open question 
\cite{schlesinger,degiorgi} concerning 
the nature of these peaks by showing that they can be interpreted as 
optical phonons. Secondly we demonstrate that the large phonon oscillator 
strength of these modes results from a coupling of the lattice
dynamical degrees of freedom to an electronic resonance above
the semiconductor gap.\cite{rice} Due to the fact that the gap closes at
high temperatures, the FeSi system provides a unique opportunity to study
this kind of coupling in detail while the gap can be swept continuously
through the phonon frequencies. Our results indicate that the
semiconductor gap has a strongly covalent character. 

We investigated the optical reflectivity of FeSi single crystals (approx.  
$2\times7$ mm$^{2}$ and thickness 1 mm) grown by the floating 
zone method. The measurements were performed in near normal incidence 
configuration over a frequency range going from 30 cm$^{-1}$ to $20\,000$ 
cm$^{-1}$, using Fourier transform spectrometers. 
The samples were mounted inside a liquid 
helium flow cryostat with temperatures ranging from 4 K to 300 K. The 
reflectivities were calibrated against a gold mirror from low frequency to 
$15000$ cm$^{-1}$ and an aluminum mirror for higher frequencies.

In Fig.\,1 the reflectivity data (R), the real part of the dielectric function 
($\epsilon_{1}$) and the dynamical conductivity ($\sigma_{1}$) 
obtained by Kramers-Kronig transform are shown up to $10\,000$ cm$^{-1}$ 
for five different temperatures. There is a strong temperature dependence in 
the data up to $6\,000$ cm$^{-1}$. Using the f-sum rule we estimate that
at room temperature the free carrier contribution corresponds to a charge 
carrier density of the order of $10^{21}$ to $10^{22}$ cm$^{-3}$ with a life-time 
of 5 fs or less, which is rather untypical for clean semiconductors. The
free carrier contribution drops rapidly upon reducing the temperature.  
This results in a strong depletion of the optical conductivity at low 
frequencies and in the onset of absorption at 570 cm$^{-1}$ (70 meV)
which we identify as a semiconducting gap in the particle-hole continuum, thus
confirming earlier reports on the infrared properties.\cite{schlesinger,degiorgi}

Below the gap we observe a number of prominent 
absorption lines which were previously attributed to phonons \cite{schlesinger} 
and partly to excitons. \cite{degiorgi} 
FeSi has a cubic structure B20 (space-group P$_{2_{1}3}$, factor-group T(23) 
and site group C$_{3}$), with four Si and four Fe atoms per unit cell. 
Using the correlation method\cite{infrared} we found for the irreducible 
representation of  the FeSi optical vibrations $\Gamma$=2A+2E+5T. From the 
factor group analysis it is possible to conclude that all these nine modes 
are Raman-active. Contrary to what was reported in an earlier 
investigation\cite{degiorgi} where only 3 infrared-active phonons were anticipated, 
in principle ({\em i.e.,} based on symmetry selection rules) 
each of the 5 triply degenerate T symmetry modes can be infrared-active.
The A and E modes are infrared-forbidden. 
Using  Raman-spectroscopy Nyhus {\em et al.} \cite{nyhus} identified 
five T modes at 193, 260, 311, 333 and 436 cm$^{-1}$, two E modes at 180 and 
315 cm$^{-1}$ and one A mode at 219 cm$^{-1}$.
In our far infrared spectra (Fig.\,1) four of these
peaks are clearly visible with resonance frequencies for the 300 K data at 198, 
318, 338 and 445 cm$^{-1}$, while the 260 cm$^{-1}$ mode apparently
has a very small optical oscillator strength and is absent in our spectra. 
So far the 338 cm$^{-1}$ mode had escaped detection in infrared 
spectroscopy,\cite{schlesinger,degiorgi} possibly due to inhomogeneous 
broadening of the lines. There is no evidence 
in our data for an optical phonon at 140\,\,\cite{degiorgi} or at 
820 cm$^{-1}$.\, \cite{schlesinger} Hence we conclude that all four
strong absorption lines can be identified as optical phonons, while
the oscillator strength of the optical phonon expected at 260 cm$^{-1}$
is very small or zero.

At low temperature the value $\epsilon_{\infty} \approx 200$ ({\em i.e.,} the 
dielectric constant at frequencies just above the last phonon 
resonance)  indicates a high 
electronic polarizability for FeSi, mainly due to virtual transitions in a
band from 700 to 1500 cm$^{-1}$, which becomes progressively masked by the 
negative free carrier contribution upon increasing the 
temperature. We will discuss 
below that the optical phonons in the far infrared region derive their 
optical oscillator strength from a coupling of the atomic coordinates to 
these charge degrees of freedom.  

Previously the high intensity of these phonon lines has been indicated as 
evidence for a strong ionicity in this compound.\cite{schlesinger} 
The transverse effective charge $e_T^*$ of the ions\cite{lucovsky} 
can be calculated directly by applying the $f$-sum 
rule to the vibrational component of $\sigma(\omega)$: 
\begin{equation}
8 \int_0^{\infty}\sigma_1(\omega') d\omega' = 
4\pi \mu^{-1}n_i e_T^{* 2} =
 \sum_{j} S_{j} \omega^2_j
\end{equation}
where $\mu$ is the reduced mass of an Fe-Si pair, $n_i$ is the number of 
Fe-Si pairs per unit volume ({\em i.e.,} 4 per unit cell), 
$S_j$ is the phonon strength,   
$\omega_j$ is the resonant frequency  and $j$ is  the index 
identifying the $j$-th phonon. 
Throughout this paper we will use the dimensionless number 
$Z=e_T^*/e$ to indicate the total transverse effective charge summed
over all phonons. In the inset of  
Fig.\,1  we display for T=4 K the quantity 
$Z^2(\omega)$, obtained by integrating $\sigma_{1}$ in Eq.\,1 up to a frequency
$\omega$. We observe, superimposed on a smooth
remnant electronic background, four steps 
$\Delta Z^2$, one for each optical phonon. Adding up these steps 
we estimate for the total transverse effective charge at 
low temperature a value $Z^2 \approx 16$. However, due to the fact 
that Fe and Si have practically the same electronegativity 
and electron affinity, strong ionicity is not expected from a chemical 
point of view, nor has it been inferred from {\em ab initio} calculations
of the electron density distribution.\cite{mattheiss,degroot} This
situation is reminiscent of the IV-VI
narrow-gap semiconductors PbS, PbSe, PbTe and SnTe where $Z$ 
values between 5 and 8 were found. For the latter systems this has
been attributed to a resonance (or mesomeric) bonding 
effect.\cite{lucovsky}  With this model the empirically found relation 
$Z^2\sim(\epsilon_{\infty}-1)$ was explained by the resonance
aspect of the chemical bond\cite{lucovsky,pauling} in these compounds. Based on
a similar concept a theory of infrared/optical
spectra of coupled vibrational and charge transfer excitations
was developed by Rice, Str\"assler and Lipari\cite{rice} in the
context of dimerized organic linear-chain conductors. 

The next step is the analysis of the phonon parameters for the different 
modes. Some of them exhibit an asymmetric line shape, probably deriving from 
an interaction between lattice vibrations and the electronic background. 
We were successful in fitting the phonon lines with
a Fano profile. Fano theory \cite{fano} 
describes the interaction of one or more discrete levels with a continuum 
of states resulting in asymmetric optical absorption peaks. 
After subtraction of the continuous electronic background the Fano 
profile is given by:
 \begin{equation}
 \sigma(\omega)=i\sigma_{0}(q-i)^2 \cdot (i+x)^{-1}
 \end{equation}
where $\sigma_{0}$ is the background, 
$x=(\omega^{2}-\omega_T^{2}) / \gamma\omega$ ($\gamma$ 
and $\omega_T$ are the line-width and the resonant frequency of 
the unperturbed vibrational state) and $q=-1/\tan(\Theta/2)$ is 
the dimensionless Fano parameter reflecting the degree of asymmetry of 
the peak (for $\Theta=0$ or, equivalently, $|q|\rightarrow \infty$ a 
Lorentz line shape is recovered). The following relation exists
between the Fano parameters and the oscillator strength:
 $ S= 4\pi \sigma_{0}(q^2-1)\gamma \omega_T^{-2} $.
The Fano parameters obtained for all the peaks are summarized in 
Fig.\,2, where we have plotted the phonon frequency shift
$\Delta \omega (T)$, 
the line-width $\gamma$, the phonon strength {\em S} and the asymmetry 
parameter $\Theta$. We can see that $\Delta \omega (T)$ and $\gamma$ 
show a very strong temperature dependence, in particular for T$>$100 K. 
Clearly there is a strong decrease of the line-width $\gamma$ upon 
reducing the temperature. This dramatic loss 
of electronic relaxational channels, which comes with the opening of
a gap in the electronic spectrum, suggests the presence of a coupling 
between electronic states and the lattice vibrations in FeSi.

The behavior of the parameter $\Theta$ is also very important   
in establishing if such a coupling exists and whether it reflects the 
temperature dependence of the gap itself. While the 338 cm$^{-1}$ line 
shows a slight asymmetry only at low temperature and  this 
mode is well described by a classical Lorentz oscillator
($|\Theta| \approx 0$) for T$>$100 K, the situation is rather different for the other 
phonons. All of them are characterized by considerable asymmetry, 
especially the one at 445 cm$^{-1}$. Significant in this analysis 
is also the sign of the parameter $\Theta$. For the lines at 198 
and 318 cm$^{-1}$ we found $\Theta>0$ at all temperatures, indicating 
a predominant interaction between these modes and electronic states  
higher in energy. On the other hand the 445 cm$^{-1}$ peak exhibits a 
negative value  of  $\Theta$ (interaction with electronic states 
lower in energy), with a maximal degree of asymmetry (maximum in $|\Theta|$) 
at T=150 K. One has to notice (Fig.\,1) that it is approximately at this
temperature that the gap edge crosses the phonon resonance frequency 
of 445 cm$^{-1}$. Moreover, for T$>$150 K the asymmetry of this mode 
changes sign, $ i.e.$,  $\Theta>0$. At T=300 K, having the gap 
already closed, the value of $\Theta$ is the same for all the 
three phonons characterized by a considerable asymmetry.

These  differences in the behavior of $\Theta$  can possibly  be 
related to the way the gap disappears (Fig.\,1). Increasing
the temperature the conductivity increases within the gap  
as a result of two different processes. First the 'closing in' of 
the gap: the movement of the gap edge from high to low frequency. 
Secondly the 'filling in' of the gap: the increase with temperature 
of the background conductivity within the gap.  Probably the 445 
cm$^{-1}$ phonon, as it is the closest one to the gap edge, is more 
strongly influenced by the 'closing in' than by the 'filling in'.  In fact, as 
we pointed out, the asymmetry of the mode at  
445 cm$^{-1}$ changes sign when the gap edge has moved below the 
resonant frequency. The other phonons, whose $\Theta$ does not show 
such a strong temperature dependence, seem to be influenced predominantly
by the 'filling in' of the gap, a process smoother than the movement 
of the gap edge.

By summing over the contributions of the four phonons, using the 
values for $S_j$ and $\omega_j$ displayed in Fig.\,2, the
temperature dependence of $e_T^*/e$ was calculated. 
The total transverse charge shown in Fig.\,2 increases with rising 
temperature and saturates at a value of $Z \approx4.6$ for T$>$150 K.

This value for $Z$ which, as mentioned, cannot be due to a real 
ionicity of the system, can be explained if we assume a coupling
between the vibrational degrees of freedom and electron-hole 
excitations across the gap of 70 meV.  These  charge excitations 
 have a large contribution of E+T character and a weak contribution of
A character, as measured in Raman-spectroscopy. \cite{nyhus} Because 
of the lack of inversion symmetry for this crystal structure there are no 
charge or vibrational excitations of pure even/odd character. 
Therefore a coupling is allowed between all five 
infrared-active T modes and the charge excitations  of  T character. As 
group theory does not provide a qualitative argument why one infrared mode is 
missing in our experimental spectra, we are led to the quantitative conclusion 
that its oscillator strength is below our detection limit.

Within this approach it is possible, considering a linear coupling between 
lattice vibrations and electronic oscillators, to account for the strong 
oscillator strength observed for the infrared-active phonons. 
For a phonon coupled to an electronic resonance 
the optical conductivity is: \cite{rice,riceprivate}
 \begin{equation}
  \sigma_j(\omega)=
  \frac{-i\omega n_e \mu^{-1} e_{T,j}^{* 2}} 
     {(1-\lambda_j)\omega_j^2-\omega(\omega+i\gamma_j)}
 \end{equation}
Here $\lambda_j = g^2 m_e^{-1}\mu^{-1}\omega_j^{-2}\omega_{p,e}^{-2}
\epsilon_e$ is a dimensionless electron-phonon 
coupling parameter and
$\epsilon_e = 
 \omega_{p,e}^2\left[\omega^2_e-\omega(\omega+i\gamma_e)\right]^{-1}$ is
the contribution to the dielectric function due to an electronic oscillator 
coupled to the phonons. The indices {\em e} and {\em j} refer to the electronic 
oscillator and the $j$-th phonon respectively. 
The transverse effective charge in this model is:
 \begin{equation}
  e_{T,j}^{* 2} = 
  \lambda_j \epsilon_en_e^{-1}n_i (Z_i e)^2 \omega_j^2 \Omega_{ph}^{-2}
  \label{etran}
 \end{equation}
The quantity $\Omega_{ph}^2=4\pi n_i\mu^{-1}Z_i^2e^2$ 
is the square plasma frequency of the lattice, where $Z_i$ is the formal 
valence of the ions. Using these experimental quantities of FeSi 
we obtain $\Omega_{ph}/(2\pi c Z_i)=341  \mbox{cm}^{-1}$. 
From a detailed fit of our data to the generalization of Eq.\,3
to several phonons coupled to the charge, which reproduces the intensities
and the Fano-line shapes of the four phonons, we obtained 
the coupling constants $\lambda_j=$ 0.063, 0.080, 0.096 and 0.036 
for the four phonons at 198, 318, 338 and 445 cm$^{-1}$ respectively at 4 K.
The Fano asymmetry parameter $\Theta$ in this model is 
$\Theta_j=2\mbox{Arg}\left[\epsilon_{e}(\omega_j)\right]$. 
To reproduce the positive and negative values of $\Theta$ for all 
phonons, we had to introduce in the model an electronic oscillator 
with a resonant frequency $\omega_{e}= 900$ cm$^{-1}$ as well as a Drude peak, 
both of them coupled to the phonons. If we
insert these values of $\lambda_j$ in Eq.\,\ref{etran},  
using the transverse effective charges of the inset of Fig.\,1, we
obtain $\epsilon_e \approx 80$ for the electronic resonance
coupled to the phonons. This implies that about $1/3$ of
the oscillator strength associated with the high electronic dielectric constant
($\epsilon_{\infty} \approx 200$) is coupled to the phonons.  
We conclude from this analysis, that the high
transverse effective charge results from a moderate coupling of the 
optical phonons to an electronic resonance with a large oscillator strength. 

An important question concerns the relation of this resonance 
to the unusual magnetic properties in FeSi. Each 
Si atom is coordinated by six Si neighbors at a distance of 2.78 $\AA$, 
while the transition-metal atom has seven Si neighbors with distances varying 
from 2.27 to 2.52 $\AA$. As the same 
structure is formed for CrSi, MnSi, FeSi, CoSi, NiSi, RuSi, RhSi, ReSi and OsSi it appears that 
the backbone of these compounds is formed by the Si and transition-metal  
outer shells. The localized $3d$ electrons on the  transition-metal 
sites  would have a magnetic moment ({\em e.g.,} $S=1$ for Fe\cite{pauling}). 
These local moments can then be compensated by the conduction electrons of the 
backbone. The many-body ground state built from a
superposition of such singlets centered at every Fe site then corresponds to a
state which could be described as a Kondo-insulator.\cite{aeppli} 
Alternatively one can postulate 
that an even number of 3d electrons resides on every Fe atom  
in a singlet ground state,\cite{tmrice}  
while thermal excitations to a high spin ({\em e.g.,} triplet) state would
be responsible for the Curie-like magnetic susceptibility at elevated temperatures. 
However, neither of these two scenario's results in a resonance hybrid\cite{pauling}
of bonds between Fe and Si. Hence our experimental evidence for such 
resonant behavior may indicate that a different theoretical 
approach is required.  

In conclusion we have investigated the optical conductivity of FeSi. We observed
the opening of a gap of the order of 70 meV upon reducing the 
temperature below 100 K. On the basis of a group theoretical analysis all 
the sharp absorption peaks observed in the gap region have been assigned
to particular lattice excitations. The position, oscillator strength and asymmetry
of the phonons were analyzed using the Fano model. The large value of 
the transverse effective charge ($e_T^* \approx4e$) along with the observed
resonance behavior of the phonon parameters 
when the gap sweeps through the phonon-frequency 
provides strong evidence for a moderate coupling ($\lambda\approx 0.1$)
between the vibrational degrees of freedom and low energy electron-hole
excitations in this compound.        

We gratefully acknowledge D. I. Khomskii, and G. A. Sawatzky 
for stimulating discussions. We thank M. J. Rice and
R. A. de Groot for theoretical support and communicating unpublished
results to us. This investigation was supported by the Netherlands Foundation for
Fundamental Research on Matter (FOM) with financial aid from
the Nederlandse Organisatie voor Wetenschappelijk Onderzoek (NWO).

\begin{figure}
\caption{Normal incidence reflectivity, real part of the dielectric function and
optical conductivity of FeSi. In the inset we present an enlarged view of 
$\sigma_{1}(\omega)$ (dotted curve) and the function 
$Z^2(\omega)=8\mu (4\pi n_i e^2)^{-1} \int_0^{\omega}\sigma(\omega')d\omega'$ 
for T=4 K.} 
\label{fig1}
\end{figure}
\begin{figure}
\caption{Temperature dependence of the frequency shift 
($\Delta\omega_j(T)=\omega_j(T) -\omega_j(300)$), 
decay rate ($\gamma_j$), Fano asymmetry parameter ($\Theta_j$) and oscillator
strength ($S_j$) of the 4 transverse optical phonons in FeSi. In the lowest section
of the right-hand panel the transverse effective charge ($Z$) summed 
over all phonons is displayed.}
\label{fig2}
\end{figure}

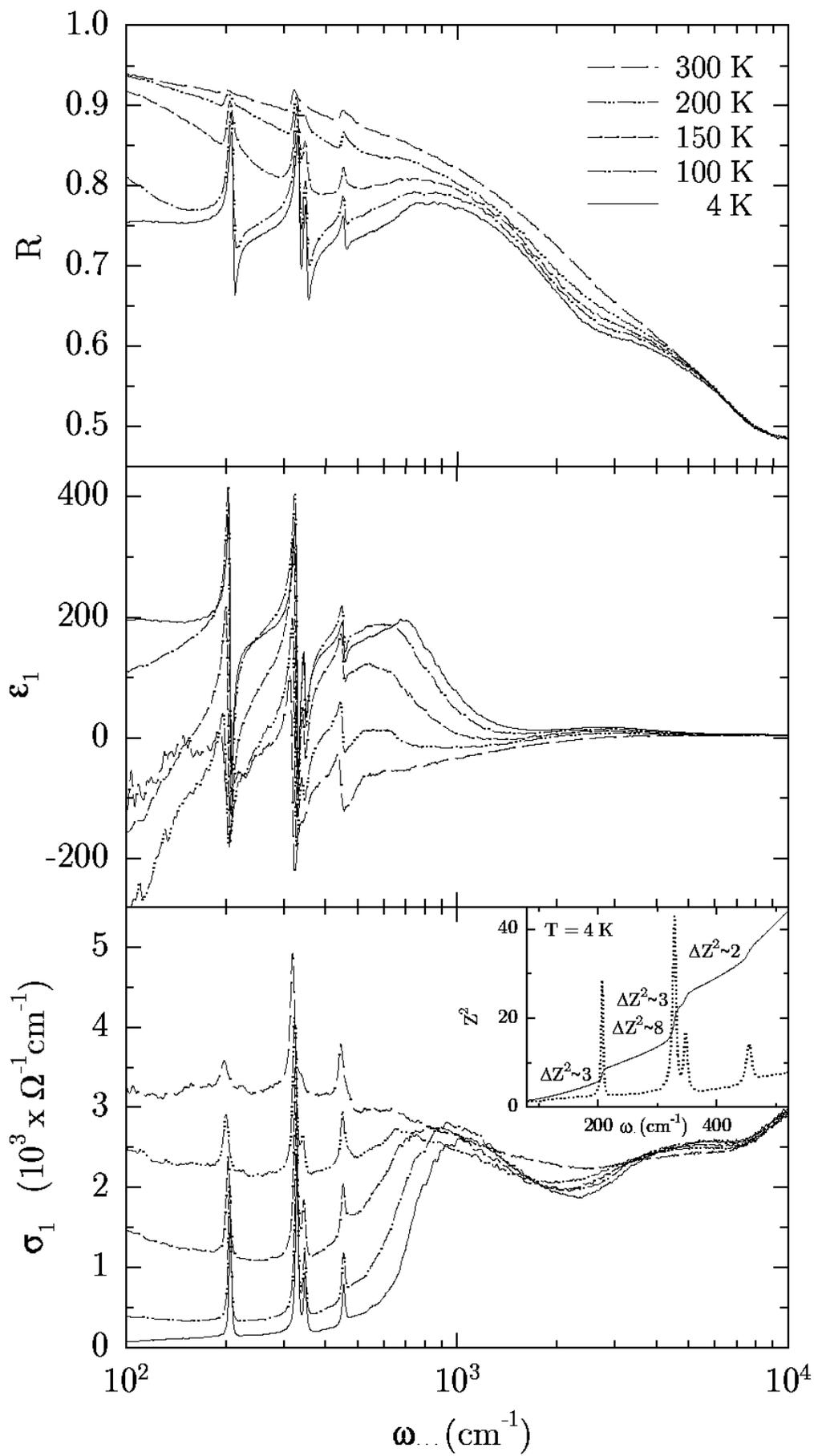
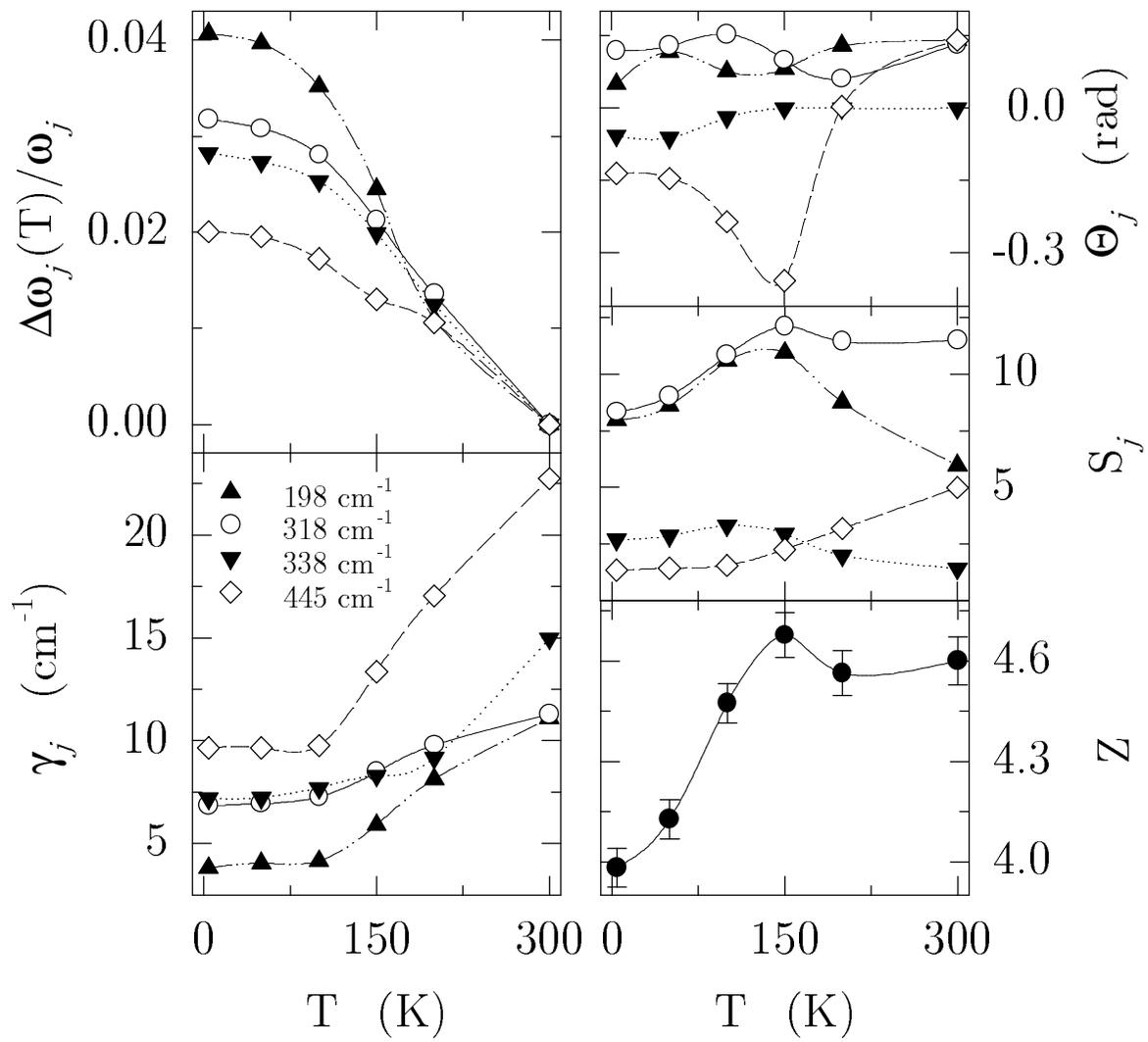
\end{document}